
\magnification=1200
\baselineskip=20pt
\hfil{NHCU-HEP-94-13}
\centerline{\bf Classical N=2 W-superalgebras From}
\centerline{\bf Superpseudodifferential Operators}
\vskip 1.5cm
\centerline{Wen-Jui Huang$^1$, J.C. Shaw$^2$ and H.C. Yen$^1$}
\vskip 1 cm
\centerline{$^1$Department of Physics, National Tsing Hua University}
\centerline{Hsinchu, Taiwan R.O.C.}
\vskip .5cm
\centerline{$^2$Department of Applied Mathematics, National Chiao Tung
University}
\centerline{Hsinchu, Taiwan R.O.C.}
\vskip 1.5cm
\centerline{\bf Abstract}
\vskip 1cm

We study the supersymmetric Gelfand-Dickey algebras associated with the
superpseudodifferential operators of positive as well as negative leading
order. We show that, upon the usual constraint, these algebras contain the
N=2 super Virasoro algebra as a subalgebra when the leading order is odd.
The decompositions of the coefficient functions into N=1 primary fields are
then obtained by covariantizing the superpseudodifferential operators.
We discuss the	problem of identifying $N=2$ supermultiplets and work out
a couple of supermultiplets by explicit computations.

\vskip 1.5cm
\noindent{PACS: 03.40.-t, 11.30.Pb}
\vfil\eject

\noindent{\bf I. Introduction}

The relevance of the study of W-algebras in two-dimensional conformal field
theory is now quite clear. The quantum W-algebras were first introduced by
Zamolodchikov as extensions of the conformal symmetry[1]. Soon after this work
it was realized that the classical $W_n$-algebras arise quite naturally as the
exotic hamiltonian structures for the generalized KdV hierarchies[2-7]. These
hamiltonian structures can be elegantly expressed by the second Gelfand-Dickey
bracket defined by differential operators[8-10]. Extensions of the
Gelfand-Dickey
bracket for pseudodifferential operators give a class of W-type algebras
called $W^{(n)}_{KP}$ which are the hamiltonian structures of the KP hierarchy
[11-15].
Recently, the supersymmetric version of the second Gelfand-Dickey brackets were
constructed[16-18]. A series of N=1 and N=2 W-superalgebras have been obtained
from the brackets defined by superdifferential operators.

In this paper, we like to study the superalgebras arising from the second
Gelfand-Dickey brackets defined by superpseudodifferential operators[19-22].
These
superalgebras, to our knowledge, are still unexplored. Our main
motivation comes from the fact that in the bosonic case all hitherto known
$W_{\infty}$-type algebras can be obtained from $W^{(n)}_{KP}$ and its
``analytic contiuation'' $W^{(q)}_{KP}$[23] via reductions, contractions  or
truncations[24]. Thus, we believe that	the superalgebras from
superpseudodifferential operators could  possibly lead to an interesting
super version of $W_{\infty}$-type algebras. The first aim of this paper is
therefore to find the $N=2$ analogue of $W^{(n)}_{KP}$. To this purpose, we
consider the usual reduction of these superalgebras. We find that it is
possible
when the leading order is a (positive or negative) odd integer. In other
words, in this case  these superalgebras contain the $N=2$ super Virasoro
algebra as a subalgebra. In order to see whether these superalgebras are
genuine $N=2$ W-superalgebras or not we need to identify the required $N=2$
supermultiplets. This is a very difficult task. We know that  in the
cases of ordinary
(pseudo)differential operators	the desired primary fields can be easily
obtained by putting the operators into a conformally covariant form[25,26].
But the superconformally covariant form of superdifferential operators can only
give us the decompositions of coefficient functions into $N=1$ primary fields
due to the fact that these super operators are defined on $(1|1)$
superspace[27,28].
The $N=2$ supermultiplets can be identified only if we further compute the
hamiltonian flow defined by the spin-1 current and redefine the $N=1$ primary
fields properly. The last step is where the difficulty lies since there is yet
no systematical way to handle the spin-1 flow. Therefore these $N=2$
supermultiplets have never been completely identified. Despite of this, we
still carry out the superconformal covariantization program for the
superpseudodifferential operators to get the series of $N=1$ primary fields.
Then we discuss the problem of identifying $N=2$ supermultiplets. In fact, we
show that the identification problem for the case of leading order $2m+1$ is
equivalent to that for the case of leading order $-2m-1$.  Moreover, two
supermultiplets are identified by explicit computations.

We organize this paper as follows. In Sec.II. we introduce the second
Gelfand-Dickey bracket for superpseudodifferential operators and show
that a reduction yields $N=2$ super Virasoro algebra if the leading order
is odd. In Sec. III. we prove that the action of superconformal transformation
on the superpseudodifferential operator is nothing but a hamiltonian flow
defined by the second Gelfand-Dickey bracket. The superconformally covariant
form of the superpseudodifferential operators is obtained. In Sec. IV. we
identify first two $N=2$ supermultiplets of the negative part of an odd-order
superpseudodifferential operator.  We present our concluding remarks in Sec.
V..

\vskip 1cm

\noindent{\bf II. Superpseudodifferential Operators and Second Gelfand-Dickey
Bracket}

We consider the superdifferential operators on a $(1|1)$ superspace with
coordinate $(x, \theta)$.  These operators are polynomials in the
supercovariant derivative $D=\partial_{\theta} + \theta \partial_x$ whose
coefficients are $N=1$ superfields;i.e.
$$ L= D^n + U_1 D^{n-1} + U_2 D^{n-2} + \dots +U_n + U_{n+1} D^{-1} + \dots
\eqno(2.1)$$
where $n$ is a nonzero integer (could be negative).
As usual, we assume that they are homogeneous under the usual $Z_2$ grading;
that is, $|U_i|=i$(mod 2). The bracket will involve functional of the form
$$ F[U] = \int_B f(U) \eqno(2.2)$$
where $f(U)$ is a homogeneous (under $Z_2$ grading) differential polynomial of
the $U_i$'s and $\int_B = \int dx d\theta$ is the Berezin integral which is
defined in the usual way, namely, if we write $U_i=u_i + \theta v_i$ and
$f(U)=a(u,v)+\theta b(u,v)$ then $\int_B f(U) = \int dx b(u,v)$. The
multiplication is given by the super Leibnitz rule:
$$ D^k \Phi = \sum^{\infty}_{i=0} \left[ \matrix{k \cr k-i} \right]
(-1)^{|\Phi|(k-i)} \Phi^{[i]} D^{k-i}, \eqno(2.3)$$
where $k$ is an arbitrary integer and $\Phi^{[i]} = (D^i \Phi)$
and the superbinomial coefficients $\left[ \matrix{k \cr i} \right]$ are
defined by
$$\left[ \matrix{k \cr k-i} \right] = \left\{ \eqalign{ & 0 \qquad for \quad
i<0 \quad or \quad (k,i) \equiv (0,1) \quad (mod \quad 2) \cr
& \pmatrix{ [{k \over 2}] \cr [{{k-i} \over 2}] }  \qquad otherwise \cr}
\right\} \eqno(2.4)$$
where $\pmatrix{p \cr q}$ is the ordinary binomial coefficient.
Next, we introduce the notions of superresidue and supertrace. Given a
super-pseudodifferential operator $P=\sum p_i D^i$ we define its superresidue
$ sresP=p_{-1} $
and its supertrace as
$StrP = \int_B sresP$
In the usual manner it can be shown that the supertrace of a supercommutator
vanishes;i.e.
$Str[P,Q] = 0$, where
$ [P,Q] \equiv PQ-(-1)^{|P||Q|}QP$
Finally, for a given functional $F[U]=\int_B f(U)$ we define its gradient $dF$
by
$$ dF=\sum^n_{k=1} (-1)^{n+k} D^{-n+k-1} {\delta f \over \delta U_k},
\eqno(2.5)$$
where
$$ {\delta f \over \delta U_k} = \sum^{\infty}_{i=0} (-1)^{|U_k|i+i(i+1)/2}
D^i {\partial f \over \partial U_k^{[i]}}. \eqno(2.6)$$
Equipped with these notions we now define the supersymmetric second
Gelfand-Dickey bracket as
$$\{F,G\}=(-1)^{|F|+|G|+n} Str[L(dFL)_+dG - (LdF)_+LdG] \eqno(2.7)$$
where $( )_+$ denotes the differential part of a super-pseudodifferential
operator. It has been shown that (2.7) indeed defines a hamiltonian structure:
it is antisupersymmetric and satisfies the super-Jacobi identity[20-22].

When $n$ is positive and when $U_{n+1}=U_{n+2}=\dots=0$ (i.e. when L is a
superdifferential operator) it can be shown that
when the constraint $U_1=0$ is imposed the
induced bracket is well-defined only when $n$ is odd[17]. The reason is that
this
constraint is second class when $n$ is odd, while becomes first class for even
$n$'s. To compute these induced brackets, we need to modify at least one of
$dF$ and $dG$ defined by (2.5) due to absence of $U_1$. The prescription is
to add a term $D^{-n} V$ to, say, dG in such a way that
$$ sres[L,D^{-n} V + dG] =0 \eqno(2.8)$$
We shall denote $X_G = D^{-n} V + dG$ for this choice of $V$. Replacing $dG$
in (2.7) by $X_G$ then gives the induced bracket. It has been shown that if we
define (of course, only when $n \ge 3$)
$$\eqalign{ T &= U_3 - {1 \over 2} U'_2 \cr
	    J &= U_2 \cr}\eqno(2.9)$$
where $V'=(DV), V''=(D^2V), \dots$ etc, then $T$ and $J$ obey the $N=2$
super Virasoro algebra:
$$\eqalign{ \{T(X), T(Y)\} &= [{1 \over 4}m(m+1)D^5 + {3 \over 2} T(X) D^2 +
{1 \over 2}T'(X)D + T''(X)] \delta(X-Y), \cr
\{T(X),J(Y)\} &= [-J(X)D^2 + {1 \over 2}J'(X)D -{1 \over 2}J''(X)] \delta(X-Y),
\cr \{J(X),T(Y)\} &= [J(X)D^2 -{1 \over 2}J'(X)D + J''(X)] \delta(X-Y), \cr
\{J(X),J(Y)\} &= -[m(m+1)D^3 + 2T(X)] \delta(X-Y), \cr} \eqno(2.10)$$
where we have written $n=2m+1$ and $\delta(X-Y)=\delta(x-y)(\theta-w)$.

The first natural question one can think of is whether or not the above result
remains true when the superpseudodifferential operators are used instead. By
straightforward calculations we can show that the answer is yes. In other
words, as long as $n$ is an odd integer $T$ and $J$ defined by (2.9) together
obey the $N=2$ super Virasoro algebra. What remains to be checked is if the
required $N=2$ supermultiplets can be defined as differential polynomails in
the coefficient functions $U_k$'s. To this end
we need to consider the hamiltonian flows defined by the two linear
functionals:
$$\eqalign{ G &= \int_B T\xi = \int_B (U_3 \xi + {1 \over 2} U_2 \xi') \cr
	    H &= \int_B J\zeta = \int_B U_2 \zeta \cr} \eqno(2.11)$$
where $|\xi(x,\theta)|=|\zeta(x,\theta)|=0$. We find that the transformations
of $L$ under the hamiltonian flows defined by $G$ and $H$ are
$$\eqalign{J(X_G) &\equiv (LX_G)_+ L - L (X_G L)_+ \cr
 &= [\xi D^2 + {1 \over 2}\xi'D + {(m+1) \over 2}\xi'']L -
		     L[\xi D^2 + {1 \over 2}\xi'D -{m \over 2}\xi''] \cr
J(X_H) &\equiv (LX_H)_+ L - L (X_H L)_+ \cr
 &= [-\zeta D-(m+1)\zeta']L - L[-\zeta D + m \zeta'] \cr}\eqno(2.12)$$
Since $T$ is the super Virasoro generator, $J(X_G)$ is called the super
Virasoro flow. If the expicit forms of (2.12) are known, one can read off
the correponding brackets at once by using the formula:
$$ J(X_F) = \sum^{\infty}_{k=2} (-1)^{k|F|+1} \{U_k,F \} D^{n-k} \eqno(2.13)$$

We shall prove in the next section that $J(X_G)$ in (2.12) is the infinitesimal
form of the superconformal covariance of $L$.

\vskip 1cm

\noindent{\bf III. Superconformally Covariant Form of $L$}

In this section we like to give the super Virasoro flow $J(X_G)$ a geometrical
interpretation and put $L$ into a superconformally covariant form.
We shall follow the construction established in refs.[27,28].
Let us recall that on the $(1|1)$ superspace with coordinate $X=(x,\theta)$.
the most general superdiffeomorphism has the form
$$\eqalign{\tilde{x} &= g(x) + \theta \kappa(x) \cr
\tilde{\theta} &= \chi(x) + \theta B(x) \cr} \eqno(3.1)$$
where $|g|=|B|=0$ and $|\kappa|=|\chi|=1$.
The superdiffeomrphism (3.1) is a superconformal transformation if
$$ D = (D \tilde{\theta}) \tilde{D} \eqno(3.2)$$

A function $f(X)$ is called a superconformal primary field of spin $h$ if,
under superconformal transformation, it transforms as
$$ f(\tilde{X}) = (D \tilde{\theta})^{-2h} f(X) \eqno(3.3)$$
We shall denote by $F_h$ the space of all superconformal primary fields of
spin $h$. As usual, a superpseudodifferential operator $\Delta$ is called  a
covariant operator if it maps $F_h$ to $F_l$ for some $h$ and $l$.

We like to study the covariance property of
$$ L = D^n + U_2 D^{n-2} + U_3 D^{n-3} + \dots + U_n + U_{n+1} D^{-1} +\dots
\eqno(3.4)$$
where we have set $U_1$ to be zero. Our aim is to see if some $h$ and $l$ can
be found so that under superconformal transformation $X \longrightarrow
\tilde{X}$
$$ L(\tilde{X}) = (D \tilde{\theta})^{-2l} L(X) (D \tilde{\theta})^{2h}
\eqno(3.5)$$
As in the case of superdifferential operators the constraint $U_1=0$ determines
both $h$ and $l$[27,28]. In fact, simple algebras gives (for any nonzero $n$)
$$ (\tilde{D})^n (D \tilde{\theta})^{-2h} = (D \tilde{\theta})^{-2h-n} (D^n +
A_{n-1} {{D^2 \tilde{\theta}} \over {D \tilde{\theta}}} D^{n-1}
 + \dots) \eqno(3.6)$$
where
$$ A_{n-1} = \left\{ \eqalign{& m \qquad \qquad (n=2m) \cr
& -2h-m \qquad (n=2m+1)  \cr} \right\}
\eqno(3.7)$$
Thus, $U_1=0$ can be preserved under
superconformal transformation only when
$$ n=2m+1, \qquad h=-{1 \over 2}m, \qquad l={1 \over 2} (m+1) \eqno(3.8)$$
In summary, we have the covariance condition
$$ L(\tilde{X}) = (D \tilde{\theta})^{-(m+1)} L(X) (D \tilde{\theta})^{-m}
\eqno(3.9)$$
The transformation laws for $U_k$'s are then completely determined by (3.9).
For example,   simple computations yield
the expected transformation laws of $J=U_2$ and $T=U_3-{1 \over 2}U'_2$:
$$\eqalign{J(X) &= J(\tilde{X}) (D \tilde{\theta})^2 \cr
T(X) &= T(\tilde{X}) (D \tilde{\theta})^3 +
 {1 \over 2}m(m+1) S(\tilde{X},X) \cr} \eqno(3.10)$$
where $S(\tilde{X},X)$ is the superschwarzian defined by
$$S(\tilde{X},X) ={{D^4 \tilde{\theta}} \over {D \tilde{\theta}}} -2 \big(
{{D^3
\tilde{\theta}} \over
{D \tilde{\theta}}} \big) \big({{D^2 \tilde{\theta}} \over {D \tilde{\theta}}}
\big) \eqno(3.11)$$

It is interesting to note that the ``central
charge''
$c_m = {1 \over 2}m(m+1)$ in (3.10) does not change sign under the sign change
of
the leading order $n=2m+1$: $\quad m \longrightarrow -m-1$. To understand this
point, let us consider the pair of superpseudodifferential operators:
$$ L^{\pm} = D^{\pm 2m \pm 1} + U_2^{\pm} D^{\pm 2m \pm 1 -2} + U_3^{\pm}
D^{\pm 2m \pm 1 -3} + \dots \eqno(3.12)$$
We shall take $L^-$ to be the formal inverse of $L^+$, that is,
$$ L^+ L^- =1  \eqno(3.13)$$
The most important point here is that (3.13) is invariant under superconformal
transformation (3.9). The equality (3.13) has fixed the functional relations
between $U^+_k$'s and $U^-_k$'s. In fact, expanding the left hand side of it
yields
$$\eqalign{ U^-_2 &= - U^+_2 \cr
	    U^-_3 &= U^+_3 -(U^+_2)' \cr} \eqno(3.14)$$
As a consequence,
$$\eqalign{ J^- &\equiv U^-_2 = - U^+_2 \equiv - J^+ \cr
T^- &\equiv U^-_3 - {1 \over 2} (U^-_2)' = U^+_3 - {1 \over 2} (U^+_2)' \equiv
T^+ \cr} \eqno(3.15)$$
It is clear now why the central charge remains unchange under
$n \longrightarrow -n$. We also like to point out that the brackets (2.10) is
invariant under $J \longrightarrow -J$. So the first of (3.15) would not harm
these brackets.

We like to show that the infinitesimal form of the covariance condition (3.9)
is nothing but the Hamiltonian flow $J(X_G)$ defined by (2.11) and (2.12).
First, we recall the most general infinitesimal
form of superconformal transformation:
$$\eqalign{ \tilde{x} &= x- \epsilon(x) - \theta \eta(x) \cr
  \tilde{\theta} &= \theta - {1 \over 2} \partial_x \epsilon(x) \theta
- \eta(x) \cr} \eqno(3.16)$$
where $|\epsilon|=0$ and $|\eta|=1$.
Defining $\xi(x) = {1 \over 2} \epsilon(x) + \theta \eta(x)$ we can show
by induction that for nonnegative integer $k$ [28]
$$(\tilde{D})^k = D^k + D[D^k,\xi]D + [D^k,\xi]D^2 + O(\xi^2) \eqno(3.17)$$
If one reexamines the proof for this equivalence in the case of
superdifferential operator given in ref.[28], one can easily
recognizes that (3.17) is the key formula. Therefore,  to
generalize
this proof to the present case we need only to prove the validity of (3.17)
when $k$ is a negative integer. To check the validity  we start with
$k=-1$. From $ D \tilde{\theta} = 1 - \xi''$  we have
$$\eqalign{ \tilde{D}^{-1} &= D^{-1} (D \tilde{\theta} )  \cr
     &= D^{-1} - D^{-1} \xi''  \cr
     &= D^{-1} - D^{-1} [D^2, \xi]  \cr
     &= D^{-1} - D \xi + D^{-1} \xi D^2  \cr
     &= D^{-1} + D [D^{-1}, \xi] D + [D^{-1}, \xi] D^2	\cr}$$
as desired. For $k < -1$ we can prove easily by induction.
With the validity of (3.17) for arbitrary integer $k$ the desired proof follows
mutatis mutandi the one of ref.[28]. We therefore conclude that
infinitesimal form of (3.9) is indeed the super Virasoro flow $J(X_G)$.

To covariantize the superpseudodifferential operators, we  briefly
review the necessary set-up[27,28].  First, we introduce a grassmanian odd
function $B(X)$ which transforms under superconformal transformation as
$$ B(\tilde{X}) = (D \tilde{\theta}) B(X) + {{D^2 \tilde{\theta}} \over
{D \tilde{\theta}}} \eqno(3.18)$$
We then make the following identification:
$$ T(X) = {{m(m+1)} \over 2} [D^2 B(X) - (D B(X)) B(X)] \eqno(3.19)$$
Clearly, (3.19) defines nothing when $m=0, -1$. This means that the
covariantization program used here is not applicable to these two cases.
As a matter of fact, it reflects that when the leading order is $\pm 1$
no $N=1$ primary basis can be defined. We can actually verify this claim via
the direct method of construction used in ref.[18].
One should note that different $B(X)$'s may actually define the same $T(X)$ as
long as its variation $ \delta B$ satisfies
$$ (\delta B)'' - (\delta B)' B - B' \delta B = 0  \eqno(3.20)$$
The transformation law of $B(X)$ enables us to introduce a covariant
superderivative defined by
$$ \hat{D}_{2k} \equiv D - 2k B(X) \eqno(3.21)$$
One can verify easily that $\hat{D}_{2k}$ maps from $F_k$ to $F_{k+{1 \over
2}}$. Hence the operator
$$\eqalign{ \hat{D}^{l}_{2k} &\equiv \hat{D}_{2k+l-1} \hat{D}_{2k+l-2} \dots
\hat{D}_{2k}  \qquad (l>0) \cr
&=[D-(2k+l-1)B][D-(2k+l-2)B] \dots [D-2kB] \cr} \eqno(3.22)$$
maps from $F_k$ to $F_{k+{l \over 2}}$. Obviously, we also need the inverse
operators of $ \hat{D}^{l}_{2k} \quad (l > 0)$, which are defined as
$$\eqalign{ \hat{D}^{-1}_{2k} &\equiv (\hat{D}_{2k-1})^{-1} = [D - (2k-1)
B]^{-1} \cr
  \hat{D}^{-l}_{2k} &\equiv \hat{D}^{-1}_{2k-l-1} \hat{D}^{-1}_{2k-l-2} \dots
\hat{D}^{-1}_{2k} \qquad (l >0) \cr} \eqno(3.23)$$
With these definitions we have the following formulae:
$$ \hat{D}_{2k} \delta B = - \delta B \hat{D}_{2k-1} + \triangle B
\eqno(3.24)$$
where  $\delta B$ is an {\it arbitrary}
variation and $ \triangle B \equiv D(\delta B) - B \delta B  $;
$$\eqalign{ \hat{D}_{2k+1} \hat{D}_{2k} \delta B &= \delta B \hat{D}_{2k}
\hat{D}_{2k-1} \cr
\hat{D}^{-1}_{2k-1} \hat{D}^{-1}_{2k} \delta B &= \delta B \hat{D}^{-1}_{2k-2}
\hat{D}^{-1}_{2k-1} \cr} \eqno(3.25)$$
where $\delta B$ is subjected to (3.20). By using (3.21)-(3.25) we can derive
(which were derived in refs.[27,28] only for positive $m$)
$$ \delta_B \hat{D}^{2m}_{2k} = -\delta B (m \hat{D}^{2m-1}_{2k} ) -
 \triangle B [m(2k+m-1) \hat{D}^{2m-2}_{2k}] \eqno(3.26)$$
and
$$ \delta_B \hat{D}^{2m+1}_{2k} = - \delta B [(2k+m) \hat{D}^{2m}_{2k}]
 - \triangle B [m(2k+m) \hat{D}^{2m-1}_{2k} ] \eqno(3.27)$$
Here $\delta B$ is subjected to the constraint (3.20).

We now write the covariant form of $L$
$$\eqalign{ L &= D^{2m+1} + U_2 D^{2m-1} + U_3 D^{2m-2} + \dots \cr
&= \hat{D}^{2m+1}_{-m} + \Delta^{(2m+1)}_2 (U_2,T) + \sum^{\infty}_{k=4}
\Delta^{(2m+1)}_{k} (W_k,T) \cr} \eqno(3.28)$$
where $W_k$ is a superconformal primary field of spin ${k \over 2}$ and
$$\Delta_p^{(2m+1)} (W_p,T) = \sum^{\infty}_{i=0} \alpha^{(2m+1)}_{p,i}
(\hat{D}^i_{p} W_{p}) \hat{D}^{2m+1-p-i}_{-m}  \qquad \alpha_{p,0}=1
\eqno(3.29)$$
The coefficients $\alpha^{(2m+1)}_{p,i}$'s are determined by requiring that
the right hand side of (3.29) depends on $B$ only through $T$. In other words,
they are solved from the recursion relations arising from the equations
$\delta_B \Delta^{(2m+1)}_p = 0$. But since (3.26) and (3.27) are valid for
 all integers $m$, we expect that the recursion relations obtained for
positive $m$[28] remain valid for nonpositive $m$. As a result, the formulae of
$\alpha^{(2m+1)}_{p,i}$ for positive $m$ remain valid for nonpositive $m$.
Therefore, without any calculations we have
$$ \eqalign{ \alpha^{(2m+1)}_{2p,2l} &= (-1)^l {{\pmatrix{l+p-m-1 \cr l }
\pmatrix{p+l-1 \cr l }} \over \pmatrix{2p+l-1 \cr l } } \cr
\alpha^{(2m+1)}_{2p,2l+1} &= {(-1)^l \over 2} {{\pmatrix{p+l-m-1 \cr l }
\pmatrix{p+l \cr l }} \over \pmatrix{2p+l \cr l } } \cr} \eqno(3.30)$$
and
$$ \eqalign{ \alpha^{(2m+1)}_{2q+1,2l} &= (-1)^l {{\pmatrix{q+l-m-1 \cr l }
\pmatrix{q+l \cr l}} \over \pmatrix{2q+l \cr l }} \cr
\alpha^{(2m+1)}_{2q+1,2l+1} &= (-1)^l {{m-q} \over 2q+1} {{\pmatrix{q+l-m \cr l
}
\pmatrix{q+l \cr l }} \over \pmatrix{2q+l+1 \cr l }} \cr} \eqno(3.31)$$

Substitutions of (3.29)-(3.31) back into (3.28) give the desired decompositions
of coefficient functions $U_k$'s into differential polynomials in $T$ and the
$N=1$ primary fields $W_k$'s.

We have seen in this section that the generalization of the covariantization
program established in refs.[27,28] to the case of superpseudodifferential
operators is quite straightforward. Key formulae like (3.17), (3.26)-(3.31)
remain unchanged at all.

\vskip 1cm
\noindent{\bf IV. N=2 Supermultiplets}

The existence of the $N=2$ super Virasoro algebra (2.10) leads naturally to
the conjecture that the $N=1$ primary fields $W_k$'s can be redefined in such
a way that $W_{2k}$ and $W_{2k+1}$ ($k \ge 2$)together form a $N=2$
supermultiplet;
i.e. under the spin-1 flow $J(X_H)$ defined by (2.11) they transform as
$$\eqalign{ \delta_{\zeta} W_{2k} &= 2 W_{2k+1} \zeta \cr
	\delta_{\zeta} W_{2k+1} &= -k W_{2k} \zeta'' + {1 \over 2} W_{2k}'
\zeta' - {1 \over 2} W_{2k}'' \zeta \cr} \eqno(4.1)$$
Since there is no simple way to handle this flow, it is even not clear whether
or not this conjecture holds in general for superdifferential operators.
Hence, we shall restrict ourselves to a very limited goal. We shall just
consider
the negative part of a superpseudodifferential operator of positive leading
order $2m+1$ ($m>0$) and present a general observation on this problem.

First, we observe that
$$ [J(X_H)]_{\pm} = [- \zeta D - (m+1) \zeta'] L_{\pm} - L_{\pm} [ - \zeta D
+ m \zeta'] \eqno(4.2) $$
that is, the positive part $L_+$ and the negative part $L_-$ transform
independently under spin-1 flow. Therefore, it is possible to consider only
the negative part. Secondly, since for a given $k > 1$ $U_{2m+k}$  is a
function of $T$ and $W_{2m+l}$'s ($k \ge l$) and since
$$\eqalign{
 \delta_{\zeta} T &= [- J D^2 + {1 \over 2} J' D - {1 \over 2} J'']\zeta \cr
\delta_{\zeta} J &= [m(m+1) D^3 + 2 T] \zeta \cr}
\eqno(4.3)$$
$\delta_{\zeta} W_{2m+k}$ must depend only on $J$, $T$ and $W_{2m+l}$'s ($k \ge
l$). As a result, the possible redefinition of $W_{2m+k}$ is  of the form
$$ \bar{W}_{2m+k} = W_{2m+k} + {\it f}_{2m+k} (J,W_{2m+1}, W_{2m+2}, \dots,
W_{2m+k-1}) \eqno(4.4)$$
where $f_{2m+k}$ is a differential polynomial.
For instance, based on the dimensional consideration, we have
$$\eqalign{ \bar{W}_{2m+2} &= W_{2m+2}	\qquad \qquad \bar{W}_{2m+3} = W_{2m+3}
\cr
\bar{W}_{2m+4} &= W_{2m+4} + a J W_{2m+2} \cr
\bar{W}_{2m+5} &= W_{2m+5} + b J W_{2m+3} \cr} \eqno(4.5)$$
It follws immediately from (4.5) that $W_{2m+2}$ and $W_{2m+3}$ must form a
$N=2$ supermultiplet if it exists at all. In the following we verify that this
is indeed true and determine the values of $a$ and $b$ which make
$\bar{W}_{2m+4}$ and $\bar{W}_{2m+5}$ form a $N=2$ supermultiplet.

Using (3.30), (3.31) and the following identities:
$$\eqalign{ \hat{D}^2_{2k} &= D^2 - B D -2k B' \cr
\hat{D}^3_{2k} &= D^3 - (2k+1) B D^2 - (2k+1) B' D - 2k B'' + 4k(k+1) BB' \cr}
\eqno(4.6)$$
$$\eqalign{ \hat{D}^{-1}_{-m} &= D^{-1} + (m+1) B D^{-2} - (m+1) B' D^{-3}
 - [(m+1) B'' + (m+1)^2 B'B] D^{-4} + \dots \cr
\hat{D}^{-2}_{-m} &= D^{-2} + B D^{-3} - (m+2) B' D^{-4} + \dots \cr
\hat{D}^{-3}_{-m} &= D^{-3} + (m+2) B D^{-4} + \dots \cr
\hat{D}^{-4}_{-m} &= D^{-4} + \dots \cr} \eqno(4.7)$$
we easily compute
$$\eqalign{ \Delta^{(2m+1)}_{2m+2} (W_{2m+2}, T)
&= W_{2m+2} D^{-1} + {1 \over 2} W_{2m+2}' D^{-2} - {1 \over 2} W_{2m+2}''
D^{-3} + \dots \cr
&\qquad \qquad -[ {{m+2} \over {2(2m+3)}} W_{2m+2}''' +
{{2(m+1)} \over {m(2m+3)}} T W_{2m+2} ] D^{-4} + \dots \cr
\Delta^{(2m+1)}_{2m+3} (W_{2m+3},T) &= W_{2m+3} D^{-2} - {1 \over {2m+3}}
W_{2m+3}' D^{-3} - {{m+2} \over {2m+3}} W_{2m+3}'' D^{-4} + \dots \cr
\Delta^{(2m+1)}_{2m+4} (W_{2m+4},T) &= W_{2m+4} D^{-3} + {1 \over 2} W_{2m+4}'
D^{-4} + \dots \cr
\Delta^{(2m+1)}_{2m+5} (W_{2m+5},T) &= W_{2m+5} D^{-4} + \dots \cr}
\eqno(4.8)$$
The desired decompositions then can be read off from (4.8):
$$\eqalign{ U_{2m+2} &= W_{2m+2} \cr
U_{2m+3} &= W_{2m+3} + {1 \over 2} W_{2m+2}' \cr
U_{2m+4} &= W_{2m+4} -{1 \over {2m+3}} W_{2m+3}' - {1 \over 2} W_{2m+2}'' \cr
U_{2m+5} &= W_{2m+5} + {1 \over 2} W_{2m+4}' - {{m+2} \over {2m+3}} W_{2m+3}''
-{{m+2} \over {2(2m+3)}} W_{2m+2}''' \cr
&\qquad \qquad \qquad - {{2(m+1)} \over {m(2m+3)}} T W_{2m+2}
\cr} \eqno(4.9)$$
Next, we find the spin-1 transformations of $U_{2m+2}, \dots, U_{2m+5}$:
$$\eqalign{ [J(X_H)]_- &= [ -\zeta D -(m+1) \zeta'] L_- - L_- [- \zeta D + m
\zeta'] \cr
&\equiv(\delta_{\zeta} U_{2m+2}) D^{-1} + (\delta_{\zeta} U_{2m+3}) D^{-2} +
(\delta_{\zeta} U_{2m+4}) D^{-3} + (\delta_{\zeta} U_{2m+5}) D^{-4} + \dots
\cr} \eqno(4.10)$$
where
$$\eqalign{
\delta_{\zeta} U_{2m+2} &= [ - U_{2m+2} D + 2 U_{2m+3} ] \zeta \cr
\delta_{\zeta} U_{2m+3} &= [ - (m+1) U_{2m+2} D^2 + U_{2m+3} D - U_{2m+3}' ]
\zeta \cr
\delta_{\zeta} U_{2m+4} &= [ - (m+1) U_{2m+2} D^3 - U_{2m+3} D^2 -(U_{2m+4}' -
2 U_{2m+5})] \zeta \cr
\delta_{\zeta} U_{2m+5} &= [ (m+1) U_{2m+2} D^4 + m U_{2m+3} D^3 - (m+2)
U_{2m+4} D^2 + U_{2m+5} D - U_{2m+5}' ] \zeta \cr} \eqno(4.11)$$
Combining (4.9) and (4.11) we finally get
$$\eqalign{
\delta_{\zeta} W_{2m+2} &= 2 W_{2m+3} \zeta \cr
\delta_{\zeta} W_{2m+3} &= [-(m+1) W_{2m+2} D^2 + {1 \over 2} W_{2m+2}' D
- {1 \over 2} W_{2m+2}'' ] \zeta \cr
\delta_{\zeta} W_{2m+4} &= 2 W_{2m+5} \zeta - {{2(m+1)} \over {m(2m+3)}}
W_{2m+2} [ m(m+1) D^3 + 2T ] \zeta \cr
\delta_{\zeta} W_{2m+5} &= [-(m+2) W_{2m+4} D^2 + {1 \over 2} W_{2m+4}' D -
{1 \over 2} W_{2m+4}''] \zeta \cr
&\qquad+ {{2(m+1)} \over {m(2m+3)}} W_{2m+3} [ m(m+1) D^3 + 2T ] \zeta \cr
&\qquad+ {{2(m+1)} \over {m(2m+3)}} W_{2m+2} [ - J D^2 + {1 \over 2} J' D - {1
\over 2} J'' ] \zeta \cr} \eqno(4.12)$$
As expected, $W_{2m+2}$ and $W_{2m+3}$ indeed form an $N=2$ supermultiplet,
while $\delta_{\zeta} W_{2m+4}$ and $\delta_{\zeta} W_{2m+5}$  both contain
some unwanted terms. Therefore we have to consider the redefinitions (4.5).
In fact, we find
$$ \delta_{\zeta} \bar{W}_{2m+4} = 2 \bar{W}_{2m+5} \zeta + 2 (a-b) J W_{2m+3}
\zeta + [a-{{2(m+1)} \over {m(2m+3)}}] (\delta_{\zeta} J ) W_{2m+2}
\eqno(4.13)$$
Hence, the only choice is
$$ a = b = {{2(m+1)} \over {m(2m+3)}} \eqno(4.14)$$
With this choice we verify
$$ \delta_{\zeta} \bar{W}_{2m+5} = [ -(m+2) \bar{W}_{2m+4} D^2 + {1 \over 2}
\bar{W}_{2m+4}' D - {1 \over 2} \bar{W}_{2m+4}'' ] \zeta \eqno(4.15)$$
as we wished.

We thus have identified the first two $N=2$ supermultiplets in the negative
part of $L$. It is natural to expect that all desired supermultiplets actually
exist.

Finally, we like to present an observation on this identification problem.
We shall show that if all required $N=2$
supermultiplets can be defined when the leading order is $2m+1$ ($m$ can be
either positive or negative), then they can    also be defined when the leading
order is $-2m-1$.  For definiteness we assume for a moment that $m>0$. We use
the notataions defined by (3.12) and (3.15) and impose the condition (3.13). We
have
observed in the previous section that (3.13) is invariant under superconformal
transformation. We now recast this statement by means of the super Virasoro
flows defined by the second Gelfand-Dickey bracket. Let $\delta^{\pm}_{\xi}
L^{\pm}$ denote the super Virasoro flows generated by $T^{\pm}$ via the
respective second Gelfand-Dickey bracket. Then (3.13) implies
$$\eqalign{ \delta^+_{\xi} L^- &= - L^- (\delta^+_{\xi} L^+) L^-  \cr
	&= - L^- \big[ (\xi D^2 + {1 \over 2} \xi' D + {{m+1} \over 2} \xi'')
L^+ - L^+ (\xi D^2 + {1 \over 2} \xi' D - {m \over 2} \xi'') \big] L^- \cr
&= [\xi D^2 + {1 \over 2} \xi' D + {{(-m-1)+1} \over 2} \xi''] L^- -
L^- [\xi D^2 + {1 \over 2} \xi' D -{{(-m-1)} \over 2} \xi''] \cr
&= \delta^-_{\xi} L^- \cr} \eqno(4.16)$$
The fact that $T^- =T^+$
together with (4.16) lead to the statement that
under the identification (3.13) the $N=1$ primary fields which appear in the
superconformally covariant form of $L^+$  are still primary fields even
when the second Gelfand-Dickey bracket of $L^-$ is used instead. As a
consequence,  decompositions of the coefficients $U^+_k$'s into $N=1$
primary fields immediately induce decompositions of $U^-_k$'s by the use of
(3.13). Next we consider the spin-1 flows which we shall denote by
$\delta^{\pm}_{\zeta} L^{\pm}$. Repeating the above steps yields
$$\eqalign{ \delta^+_{\zeta} L^- &= - L^- (\delta^+_{\zeta} L^+) L^- \cr
				 &=  [-\zeta D +m \zeta'] L^- - L^-[- \zeta D
-(m+1) \zeta'] \cr
&= \delta^-_{\zeta} L^- \cr} \eqno(4.17)$$
Now since $J^- = -J^+$ we conclude that  the second Gelfand-Dickey brackets
of $L^+$ and $L^-$ both lead to the same spin-1 flow (up to an overall sign)
when the functional $H^+=\int _B J^+ \zeta$ is used in either bracket.
More explicitly, what we have shown so far is that for any functional $F$:
$$\eqalign{ \{F, T^-(X) \}^- &= \{F, T^+(X) \}^+ = \{F, T^-(X) \}^+ \cr
 \{F, J^-(X) \}^- &= \{F, J^+(X) \}^+ = -\{F, J^-(X) \}^+  \cr}\eqno(4.18)$$
where $\{ ,  \}^{\pm}$ denote the second Gelfand-Dickey bracket of
$L^{\pm}$
respectively. It is clear now that if $W_{2k}$ and $W_{2k+2}$ form an $N=2$
supermultiplet with respect to $\{ ,  \}^+$ then they will also do with
respect to $\{ ,  \}^-$. Therefore, once the required $N=2$
supermultiplets have been identified for $L^+$ the corresponding task for $L^-$
is automatically done. Interchanging the roles of $L^+$ and $L^-$ obviously
give the proof for $m<0$.  This completes the proof for the above claim.

\vskip 1cm

\noindent{\bf V. Concluding Remarks}

In this paper we have discussed the $N=2$ superalgebras arising from the
second Gelfand-Dickey bracket of superpseudodifferential operators. We find
that the forms of several formulae derived previously for the case of
superdifferential remain unchanged in this case. In other words, the
generalization is pretty straightforward. For example, the formulae (3.30) and
(3.31) obtained in refs.[27,28] immediately give us the superconformally
covariant form of superpseudodifferential operators. Hence, the biggest
problem regarding the spectrum of these superalgebras is still the
identifications of $N=2$ supermultiplets. Since the positive part and the
negative part of a superpseudodifferential operator transform independently
under the super Virasoro flow as well as the spin-1 flow, unless the
identification problem can be solved for pure superdifferential the
resolution of this problem in the present case is not possible. We like to
remark that in refs.[27,29] it is observed that when $L= D^5 + U_2 D^3 +
\dots + U_5$ the $N=1$ primary fields arising from the Drinfeld-Sokolov type
matrix formalation[29,30] form precisely the desired $N=2$ supermultiplets.
One might suspect that the matrix formulation might be helpful to this
problem. Hence, it seems worthwhile to discuss the spin-1 flow in the context
of matrix formultion. Finally, we like to remark that it would be interesting
to investigate all possible reductions, contractions and trunctions of these
$W^{(n)}_{KP}$-type superalgebras. Hopefully, some interesting
$W_{\infty}$-type superalgebras can emerge. Work in this direction is in
progress.

\vskip 1.5cm
\noindent{\bf Acknowledgements}

This work was supported by the National Science Council of Republic of China
under Grant No. NSC-83-0208-007-008.

\vskip 1.5cm
\noindent{\bf References}

\item{[1]} A.B. Zamolochikov, Theor. Math. Phys. {\bf 65}, 1205 (1985).
\item{[2]} J.-L. Gervais, Phys. Lett. {\bf 160B}, 277 (1985).
\item{[3]} J.-L. Gervais and A. Neveu, Nucl. Phys. {\bf B264}, 557(1986).
\item{[4]} T.G. Khovanova, Funct. Anal. Appl. {\bf 21}, 332 (1987).
\item{[5]} P. Mathieu, Phys. Lett. {\bf 208B}, 101 (1988).
\item{[6]} I. Bakas, Nucl. Phys. {\bf 302}, 189 (1988); Phys. Lett. {\bf 213B},
 313 (1988).
\item{[7]} Q. Wang, P.K. Panigraphi, U. Sukhatme and W.-K. Keung, Nucl. Phys.
{\bf B344}, 194 (1990).
\item{[8]} I.M. Gelfand and L.A. Dickey, Funct. Anal. Appl. {\bf 11}, 93
(1977).
\item{[9]} M. Adler, Invent. Math. {\bf 50}, 219 (1979).
\item{[10]} B.A. Kuperschmidt and G. Wilson, Invent. Math. {\bf 62}, 403
(1981).
\item{[11]} L.A. Dickey, Ann. NY Acad. Sci. {\bf 491}, 131 (1987).
\item{[12]} J.M. Figueroa-O'Farrill, J. Mas and E. Ramos, Phys. Lett. {\bf
266B}, 298 (1991).
\item{[13]} A. Das and W.-J. Huang and S. Panda, Phys. Lett. {\bf 271B}, 109
(1991).
\item{[14]} A.O. Radul in: Applied methods of nonlinear analysis and control,
eds. A. Mironov, V. Moroz and M. Tshernjatin(MGU, Moscow 1987)[in Russian].
\item{[15]} A. Das and W.-J. Huang, J. Math. Phys. {\bf 33}, 2487 (1992).
\item{[16]} J.M. Figueroa-O'Farrill, J. Mas and E. Ramos, Phys. Lett. {\bf
262B}, 265 (1991).
\item{[17]} J.M. Figueroa-O'Farrill and E. Ramos, Nucl. Phys. {\bf B368}, 361
(1991).
\item{[18]} K. Huitu and D. Nemeschansky, Mod. Phys. Lett. {\bf A6}, 3179
(1991); C.M. Yung and R.C. Warner, J. Math. Phys. {\bf 34}, 4050 (1993).
\item{[19]} F. Yu, J. Math. Phys. {\bf 33}, 3180 (1992).
\item{[20]} A. Das and W.-J. Huang, Mod. Phys. Lett. {\bf A7}, 2159 (1992).
\item{[21]} C.M. Yung, Mod. Phys. Lett. {\bf A8}, 129 (1993).
\item{[22]} E. Ramos and S. Stanciu, `` On the supersymmetric BKP-hierarchy'',
preprint QMW-PH-94-3 (hep-th/9402056).
\item{[23]} J.M. Figueroa-O'Farrill, J. Mas and E. Ramos, A one-parameter
family of hamiltonian structures of the KP hierarchy and a continuous
deformation of the $W_{KP}$ algebra, preprint BONN-HE-92-20, US-FT-92/7,
KUL-TF-922/20 (hep-th/9207092).
\item{[24]} J.M. Figueroa-O'Farrill, J. Mas and E. Ramos, Phys. Lett. {\bf
299B}, 41 (1993).
\item{[25]} P. Di Francesco, C. Itzykson and J.-B. Zuber, Comm. Math. Phys.
{\bf 140}, 543 (1991).
\item{[26]} W.-J. Huang, J. Math. Phys. {\bf 35}, 993 (1994).
\item{[27]} F. Gieres and S. Theisen, J. Math. Phys. {\bf 34}, 5964 (1993).
\item{[28]} W.-J. Huang, Superconformal covariantization of superdifferential
operators on $(1|1)$ superspace and classical N=2 W-superalgebras, to appear in
J. Math. Phys. (hep-th/9310192).
\item{[29]} F. Gieres and S. Theisen, Int. J. Mod. Phys. {\bf A9}, 383 (1994).
\item{[30]} V.G. Drinfeld and V.V. Sokolov, J. Sov. Math. {\bf 30}, 1975
(1985).

\end